\documentclass[twocolumn,prb,showpacs]{revtex4}

\usepackage{graphicx}
\usepackage{dcolumn}
\usepackage{bm}
\usepackage{braket}

\begin{document}

\title{Low-temperature density matrix renormalization group using regulated polynomial expansion
}

\author{Shigetoshi Sota}
\author{Takami Tohyama}
\affiliation{
Yukawa Institute for Theoretical Physics, Kyoto University,
Kyoto 606-8502, Japan
}


\date{\today}

\begin{abstract}
We propose a density matrix renormalization group (DMRG) technique at finite temperatures. As is the case of the ground state DMRG, we use a single-target state that is calculated by making use of a regulated polynomial expansion. Both static and dynamical quantities are obtained after a random-sampling and averaging procedure. We apply this technique to the one-dimensional Hubbard model at half filling and find that this gives excellent results at low temperatures. 

\end{abstract}


\maketitle

The density matrix renormalization group (DMRG) method~\cite{White} is a powerful numerical technique to investigate various properties of low-dimensional strongly correlated electron systems. The ground state properties are accurately calculated through targeting the ground state in each step of the DMRG process. For dynamical quantities, a multitarget procedure has been proposed~\cite{Jakelmann} and provides accurate description of various excitation spectra. These successes of DMRG come from proper choices of the target state to be necessary to construct the density matrix that contains important bases for the precise description of physical quantities. 

The extension of such a targeting procedure to thermodynamic properties has been done for one-dimensional (1D) spin systems.~\cite{Moukouri} Several tens of lowest-energy eigenstates of the systems have been taken as the target states and the density matrix is constructed by weighting a Boltzmann factor for each eigenstate. Later, the transfer-matrix method~\cite{Shibata,Wang} has been introduced to calculate thermodynamic properties in the infinite-size system. The DMRG method is employed for the calculation of the maximum eigenvalue of a transfer matrix that gives the free energy of the system. Recently another finite-temperature DMRG method has been proposed as a generalization of time-dependent DMRG.~\cite{Feiguin} These finite-temperature DMRG techniques successfully give excellent results for 1D spin and electronic systems. 
However, there are some difficulties in each technique; for example, the transfer matrix method is not easily applied to complicated models since the transfer matrix is not Hermitian. In the time-evolution method, low-temperature properties are not easily obtained since long time evolution of the system is necessary. 
Therefore, it is desired to develop a variety of finite-temperature DMRG techniques, from which one can choose the best one suitable for a given model.

In this Brief Report, we propose a scheme of DMRG at finite temperatures, which is a straightforward extension of the target-state procedure at zero temperature. The target state is weighted by a Boltzmann factor. Making use of the polynomial expansion and random sampling, we can calculate static and dynamical quantities at finite temperatures. In order to obtain good convergency at high temperature, we need a large truncation number of the density matrix. The proposed method is, therefore, suitable for lower temperature region. As a demonstration of the method, we show the specific heat, spin-spin correlation function, and dynamical current-current correlation function of the 1D Hubbard model at half filling. The DMRG results reproduce the exact digitalization results at low temperature.

The DMRG procedure at zero temperature requires a target state in order to obtain the ground-state properties. Even for finite temperatures, it may be possible to have a target state suitable for the evaluation of physical quantities. A possible target state may be given by
\begin{eqnarray}
\ket{\tilde{\xi}}\equiv e^{-\beta \hat{H}/2} \ket{\xi}=\sum_{n=1}^N e^{-\beta \epsilon_n/2} a_n\ket{\epsilon_n},
\label{eq1}
\end{eqnarray}
where $\hat{H}$ is the Hamiltonian, $\ket{\xi}$ is a normalized arbitrary vector, $\beta$ is the inverse temperature $1/T$, $N$ is the dimension of the superblock, and $a_n=\braket{\epsilon_n | \xi}$, with $\ket{\epsilon_n}$ being the eigenvector corresponding to the eigenvalue $\epsilon_n$. 
The inner product of Eq.~(\ref{eq1}) gives the partition function $Z$, provided that  $a_n^2=1$: $Z=\braket{\tilde{\xi}|\tilde{\xi}}=\sum_{n=1}^N e^{-\beta\epsilon_n}$.
Therefore, Eq.~(\ref{eq1}) is a good candidate for the target in the DMRG procedure. However, it is difficult to obtain all of the eigenstates $\ket{\epsilon_n}$ for the superblock Hamiltonian whose size of the Hilbert space is of the order of $16m^2$ in the case of the single-band Hubbard model, with $m$ being the truncation number of the density matrix. We thus need to develop a different technique to treat the operator $e^{-\beta\hat{H}/2}$ precisely without obtaining $\ket{\epsilon_n}$.

By using the Legendre polynomial expansion for the delta function, i.e.,
\begin{eqnarray}
\delta(x-x^{\prime})
=\sum_{l=0}^{\infty} w^{-1}_{l} P_l(x) P_l(x^{\prime})
\label{eq2}
\end{eqnarray}
with $w_l=2/(2l+1)$, the Boltzmann factor reads
\begin{eqnarray}
e^{-\beta \tilde{E}_n}=\int_{-1}^{1} d\epsilon e^{-\beta \epsilon} \sum_{l=0}^{\infty} w_l^{-1} P_l(\epsilon) P_l(\tilde{E}_n),
\label{eq3}
\end{eqnarray}
where $\tilde{E}_n$ is an eigenvalue rescaled to be confined within the interval of $[-1,1]$. The corresponding rescaled Hamiltonian $\hat{H}_s$ is defined as $\hat{H}_s= w_H(\hat{H}-\lambda)$ with scaling parameters $w_H$ and $\lambda$.

In general, there appear so-called Gibbs oscillations in any polynomial expansion including the Chebyshev polynomial often used in the literatures.~\cite{Vorter, LWang} The oscillations can be eliminated by introducing the Gaussian distribution function for 
$P_l(\tilde{E}_n)$ in Eq.~(\ref{eq2}).~\cite{Sota}
The polynomial regulated by the Gaussian is defined as
\begin{eqnarray}
\langle P_l(\tilde{E}_n) \rangle_{\sigma}= \frac{1}{\sqrt{2\pi\sigma^2}}\int_{-1}^{1} d\epsilon e^{-\frac{(\epsilon-\tilde{E}_n)^2}{2\sigma^2}}P_l(\epsilon),
\label{eq4}
\end{eqnarray}
where $\sigma$ is the half width of the Gaussian distribution function set to be $2\pi/L$, where $L$ denotes the highest number of $l$ in the expansion. Inserting the Boltzmann factor [Eq.~(\ref{eq3})] into the target state [Eq.~(\ref{eq1})] and returning to the operator representation, we obtain 
\begin{eqnarray}
\ket{\tilde{\xi}}\simeq \int_{-1}^1 d\epsilon e^{-\beta \epsilon/2} \sum_{l=0}^{L} w_l^{-1}P_l(\epsilon)\langle P_l(\hat{H}_s) \rangle_\sigma \ket{\xi}.
\label{eq5}
\end{eqnarray}
Since the integration in Eq.~(\ref{eq5}) with respect to $\epsilon$ leads to the modified spherical Bessel function $i_l(x)$ of the first kind, the target state is finally written as
\begin{eqnarray}
\ket{\tilde{\xi}}\simeq
C(\beta) \sum_{l=0}^{L} w_l^{-1} i_l (-\beta /2) \langle P_l(\hat{H}_s) \rangle_{\sigma} \ket{\xi},
\label{eq6}
\end{eqnarray}
where $C(\beta)$ is a normalization constant. 

In order to calculate $\langle P_l(\hat{H}_s) \rangle_\sigma\ket{\xi}$, we employ a coalitional recursive relation~\cite{Sota}
\begin{eqnarray}
&&\langle P_{l+1}(\hat{H}_s) \rangle_{\sigma}\ket{\xi}=\frac{2l+1}{l+1}\hat{H}_s\langle P_l(\hat{H}_s) \rangle_{\sigma}\ket{\xi}   \nonumber\\
&&-\frac{l}{l+1} \langle P_{l-1}(\hat{H}_s) \rangle_{\sigma}\ket{\xi}
+\frac{2l+1}{l+1}\sigma^2 \langle P_l^{\prime}(\hat{H}_s) \rangle_{\sigma}\ket{\xi}, 
\label{eq8}
\end{eqnarray}
\begin{eqnarray}
\langle P_{l+1}^{\prime}(\hat{H}_s) \rangle_{\sigma}\ket{\xi}&=&(2l+1)\langle P_l(\hat{H}_s) \rangle_{\sigma}\ket{\xi} \nonumber\\
&&+\langle P^{\prime}_{l-1}(\hat{H}_s) \rangle_{\sigma}\ket{\xi},
\label{eq9}
\end{eqnarray}
where $P_l^{\prime}(\epsilon)\equiv dP_l(\epsilon)/d\epsilon$. Starting from $\langle P_0(\hat{H}_s )\rangle_\sigma \ket{\xi}=\ket{\xi}$ and $\langle P_1(\hat{H}_s) \rangle_\sigma \ket{\xi} = \hat{H}_s \ket{\xi}$, 
we recursively calculate $\langle P_l(\hat{H}_s) \rangle_\sigma \ket{\xi}$ up to $l=L$ and construct the target state in Eq.~(\ref{eq6}). 

In the DMRG procedure, physical quantities are measured when the system size is reached to a given number in the infinite-size algorithm or enough convergency is obtained in the finite-size algorithm.~\cite{White} At this stage, we need to introduce a technique to guarantee the relation $a_n^2=1$ for the coefficients in Eq.~(\ref{eq1}). This is achieved by taking the random sampling of the state $\ket{\xi}$ and averaging over the samplings. Let us represent a randomly generated $\ket{\xi}$ as $\ket{\xi}=\sum_i r_i \ket{\xi_i}$, where $\ket{\xi_i}$ is the basis state of the system and 
$r_i$ is a normalized random number generated from a rectangular distribution whose center is at zero. 
Expanding the eigenstate $\left|\epsilon_n\right>$ also in terms of $\left| \xi_i\right>$, i.e., $\left|\epsilon_n\right>=\sum_i b_{n,i}\left| \xi_i\right>$, we obtain $a_n^2=\sum_i r_i^2b_{n,i}^2+2\sum_{i\not= j}r_i r_j b_{n,i} b_{n,j}$.
After averaging over many samplings whose number is $M$, $r_i^2$ will become a constant approximately independent of $i$, and $r_ir_j$ will vanish according to $1/\sqrt{m^2M}$.~\cite{Prelovsek} 
%
Therefore, a relation $a_n^2\simeq 1\: (n=1,\cdots,N)$ is expected to be satisfied. 
%
%

Physical quantities that do not commute with $\hat{H}$ are also obtained by using the same random sampling. An expectation value of an operator $\hat{A}$ is given by
\begin{eqnarray}
\braket{\tilde{\xi}|\hat{A}|\tilde{\xi}}
&=&\sum_n a_n^2 e^{-\beta\epsilon_n}\braket{\epsilon_n|\hat{A}|\epsilon_n}
\nonumber\\
&&
+ \sum_{n\not= m} a_na_m e^{-\beta(\epsilon_n+\epsilon_m)/2}\braket{\epsilon_m|\hat{A}|\epsilon_n}.
\label{eq10}
\end{eqnarray}
The coefficient for the off-diagonal term is expressed as $a_n a_m =\sum_i r_i^2b_{n,i}b_{m,i}+\sum_{i \neq j} r_i r_j b_{n,i} b_{m,j}$.
Since $r_i^2$ is a constant and $r_ir_j$ is zero after the sample averaging, $a_na_m$ is expected to be zero. This means that Eq.~(\ref{eq10}) gives thermodynamical average of a given quantity. 
%

We can also calculate dynamical quantities at finite-temperature.  A dynamical correlation function for an operator $\hat{A}$ may be defined as
\begin{eqnarray}
\chi_A(\omega)&=&\frac{1}{Z}\sum_n e^{-\beta\epsilon_n} 
\nonumber\\
&&\times \mathrm{Im}\bra{\epsilon_n} \hat{A} \frac{1}{\omega-\hat{H}+\epsilon_n-i\gamma} \hat{A} \ket{\epsilon_n},
\label{eq12}
\end{eqnarray}
where $\gamma$ is a small positive number.
In order to obtain $\chi_A(\omega)$, we introduce a following expression:
\begin{eqnarray}
\tilde{{\chi}}_A(\omega)
&=&\frac{1}{Z}\int_{-1}^{1} d\epsilon e^{-\beta\epsilon}
\nonumber\\
&&\times\mathrm{Im} \bra{\epsilon} \hat{A} \frac{1}{\omega-\hat{H}+\epsilon_s-i\gamma} \hat{A} \ket{\epsilon}
\label{eq13}
\end{eqnarray}
with $\ket{\epsilon}= \sum_{l=0}^{L} w^{-1}_{l} P_l(\epsilon) \langle P_l(\hat{H_s})\rangle_{\sigma} \ket{\xi}$, and $\epsilon_s=\epsilon/w_H+\lambda$. If we take $L\rightarrow\infty$ and use the random averaging, we can easily find that $\tilde{\chi}_A(\omega)$ is identical to $\chi_A(\omega)$. 
We use Eq.~(\ref{eq13}) at the stage of the measurement.

Before reaching the measurement, we need to perform a dynamical DMRG procedure using a multi-targeting technique.~\cite{Jakelmann,Matsueda} 
In order to make use of $\ket{\tilde{\xi}}$ as one of the multi targets, we introduce alternative expression 
\begin{eqnarray}
\frac{1}{Z} \int_{-1}^{1} d\epsilon 
\mathrm{Im} \bra{\epsilon} e^{-\beta\epsilon/2} \hat{A} \frac{1}{\omega-\hat{H}+\epsilon_s-i\gamma} \hat{A} \ket{\tilde{\xi}},
\label{alternative}
\end{eqnarray}
which gives $\chi_A(\omega)$ after taking $L \rightarrow \infty$ and the random averaging. In addition to $\ket{\tilde{\xi}}$, possible target states may be $\hat{A} \ket{\tilde{\xi}}$ and $\int_{-1}^{1} d\epsilon [\omega-\hat{H}+\epsilon_s-i\gamma]^{-1} \hat{A} e^{-\beta\epsilon/2} \ket{\epsilon}$. However, the later state is not easily calculated since it contains the integration in terms of $\epsilon$. Instead of this state, we introduce a state with simple form $[\omega-\hat{H}+\tilde{E}-i\gamma]^{-1} \hat{A} \ket{\xi}$, by replacing $\epsilon_s$ to an $\epsilon$-independent quantity $\tilde{E}=\braket{\tilde{\xi}| \hat{H} | \tilde{\xi}}$. This replacement is based on the fact that the dominant contribution of $\epsilon_s$ comes from the energy range where the product of the density of the eingenstates and the Boltzmann distribution function is large. In spite of such a rough approximation, we will find that this works practically well as discussed below. If this replacement does not work well, we should employ the original state with the integration as a target state. 

In the present DMRG approach, there are three parameters: the number of sampling $M$, the polynomial expansion truncation number $L$, and the DMRG truncation number $m$. 
Among them, $M$ is dependent on physical quantities as will be mentioned below. 
%
%
The number of $L$ in Eq.~(\ref{eq6}) predominantly depends on temperature $T$. The lower $T$ is, the larger $L$ is. However, it is difficult to obtain numerically the several hundred order of $i_l(-\beta/2)$. In such a low-temperature (large $\beta$) region 
where large $L$ is required, 
we introduce a smaller $\beta^{\prime}$ with a relation $\beta=n\beta^{\prime}$ ($n$ is a positive integer), and then perform the polynomial expansion in Eq.~(\ref{eq6}) $n$ times  starting from $\ket{\tilde{\xi}}$ at $\beta^{\prime}$ and inputting the obtained $\ket{\tilde{\xi}}$ into $\ket{\xi}$. At low temperature, although large $L$ is required, $m$ can be reduced as compared with that for higher temperature, since the number of the basis necessary to describe low-temperature properties is small. As a result, the computing time is shorter at low temperature than at high temperature, in order to get the same level of convergence.
%


\begin{figure}[t]
\begin{center}
\includegraphics[width=90mm]{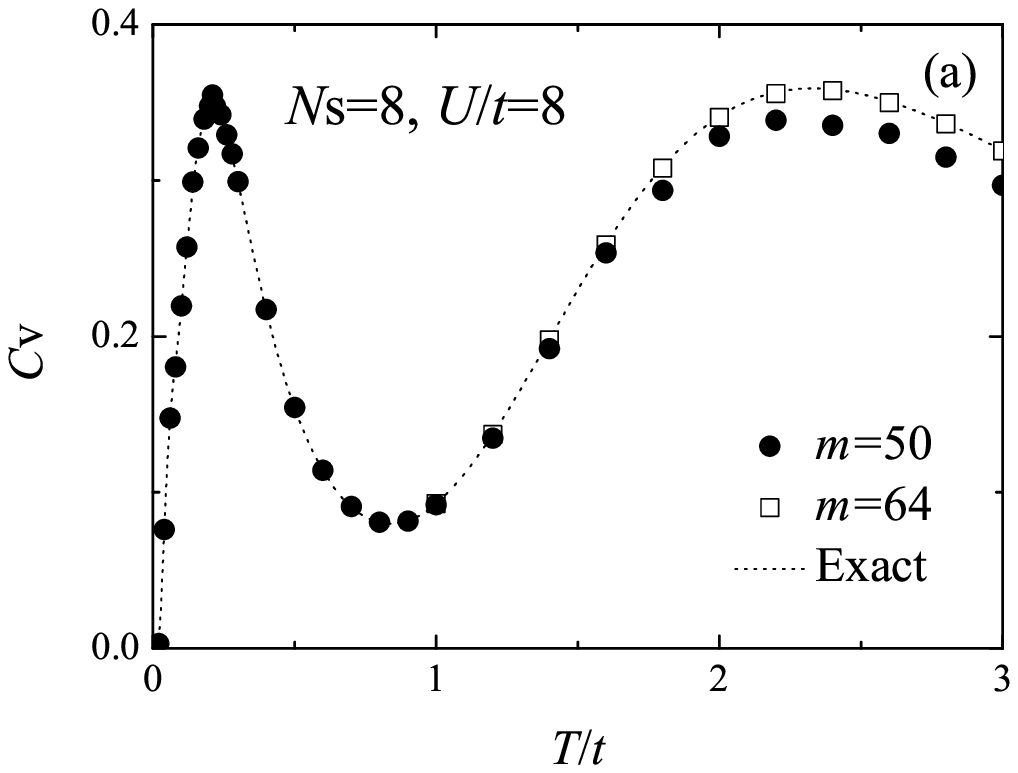}
\includegraphics[width=90mm]{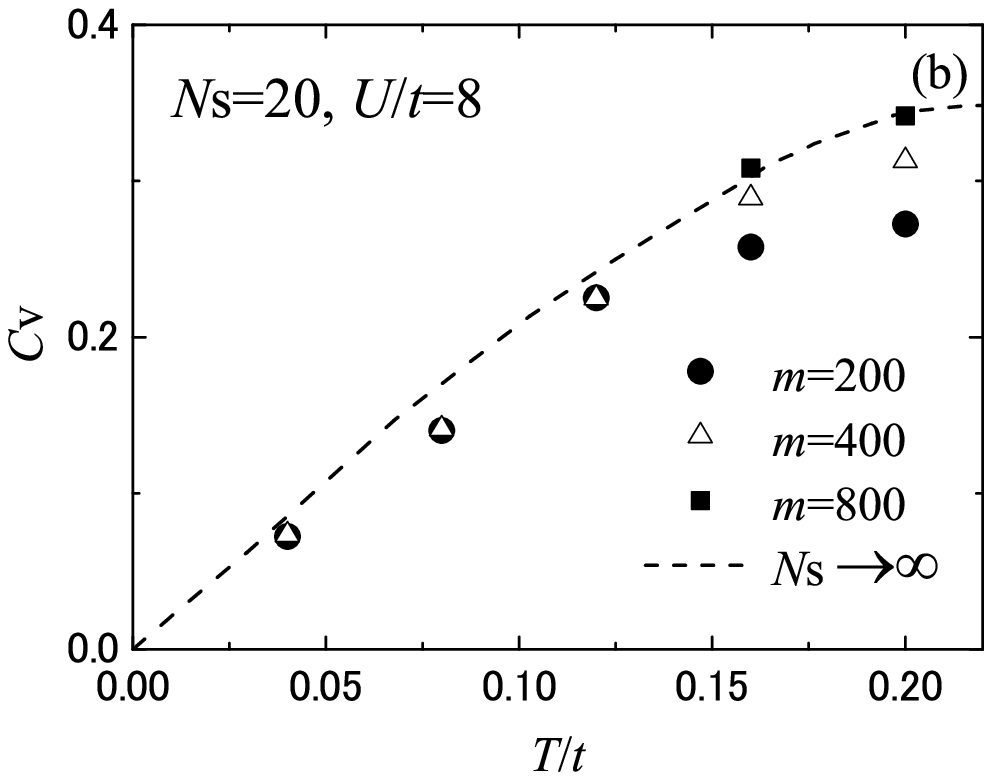}
\caption{Specific heat as a function of temperature in the 1D Hubbard model at half filling with $U/t=8$. (a) $N_\mathrm{s}=8$ and $M=1000$.  Dotted line shows the exact result obtained by the direct diagonalization. (b) $N_\mathrm{s}=20$ and $M=400$. Dash line shows the exact result in thermodynamic limit.(Ref. 12)
}
\end{center}
\label{fig1}
\end{figure}
We apply a different finite-temperature DMRG method to the 1D Hubbard model at half filling to check its efficiency. The Hamiltonian is given by
$
\hat{H}=-t\sum_{i, \sigma} 
( c_{i,\sigma}^{\dagger}c_{i+1,\sigma}+\mathrm{h.c.}
) +U\sum_i n_{i,\uparrow} n_{i,\downarrow},
$
where $c_{i,\sigma}^{\dagger}$ ($c_{i,\sigma}$) is a creation (annihilation) operator of an electron with site $i$ and spin $\sigma$, $n_{i,\sigma}=c_{i,\sigma}^{\dagger}c_{i,\sigma}$, $t$ is the hopping integral, and $U$ is the on-site Coulomb repulsion. We use a lattice with open boundary condition.
 
Figure 1 shows the specific heat obtained by using the formula $C_\mathrm{v}(T)={(N_\mathrm{s}T^2)^{-1}} ( \braket{\tilde{\xi} | \hat{H}^2 | \tilde{\xi} } - \braket{\tilde{\xi} | \hat{H} | \tilde{\xi} }^2 )$, where $N_\mathrm{s}$ is the number of sites. The result for $N_\mathrm{s}=8$ and $U/t=8$ is shown in Fig.~1(a), where the exact $C_\mathrm{v}$ is also plotted for comparison. We employ parameters of $m=50$
and 64
, $L=80$ $(T/t=0.1)$, and $M=1000$. The error bar due to the sampling is within the size of the symbols. At $m=50$, the low-temperature peak of $C_\mathrm{v}$ agrees with the exact result. However, with increasing $T$ the deviation from the exact result is enhanced. This is improved if we use larger $m$. In fact, taking $m=64$, we get complete agreement between the exact and DMRG results, since $16m^2$ exceeds the dimension of the Hilbert space. Such an improvement is also clear in the case of $N_\mathrm{s}=20$ as shown in Fig.~1(b).
Temperatures are restricted to a range just below the first peak of $C_\mathrm{v}$. The exact result in the thermodynamic limit $N_\mathrm{s}\rightarrow \infty$ (Ref. 12)
is also shown for comparison. 
%
%
We find that $m=200$ is enough to get good convergence below $T/t=0.12$. At higher temperatures, a larger $m$ is required but it is still accessible by using a standard computing system. From these results, it is apparent that this DMRG method is efficient for the calculation of low-temperature properties.

\begin{figure}[t]
\begin{center}
\includegraphics[width=90mm]{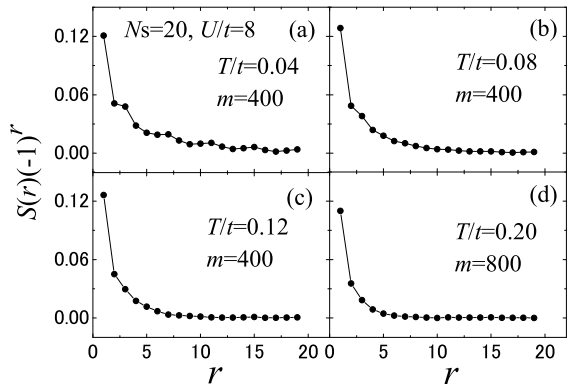}
\caption{Spin-spin correlation function in the 20-site Hubbard chain at half filling with $U/t=8$. $M=400$. }
\end{center}
\label{fig2}
\end{figure}

As a quantity whose operator does not commute with $\hat{H}$, we choose the
 spin-spin correlation function $S(r) \equiv  \braket{\tilde{\xi}|S^z_iS^z_{i+r}|\tilde{\xi}}$, where $S^z_i$ is the $z$ component of the total spin operator at site $i$. We choose the two sites, $i$ and $i+r$, in order to make the central site of a given lattice the middle of them. Figure~2 shows the staggered correlation $S(r)(-1)^r$ at several temperatures for $N_\mathrm{s}=20$ and $U/t=8$. The truncation number $m$ is changed with temperature in order to get good convergence. The spin correlation decreases with increasing temperature as expected. 

Finally we show the dynamical current-current correlation function in Fig.~3. The operator $\hat{A}$ in Eq.~(\ref{eq13}) is replaced by the current operator $\hat{j}=it\sum_{i,\sigma}( c^\dagger_{i,\sigma} c_{i+1,\sigma} - \mathrm{h. c.} )$. 
In this calculation, we employ $\gamma=0.2t$. Although $\gamma\rightarrow 0$ is desired in general, the finite value is introduced here in order to reduce the computational time, in particular, at high energy region of $\tilde{\chi_j}(\omega)$. 
We compare the DMRG results of $\tilde{\chi_j}$ with the exact one for $N_\mathrm{s}=8$. At $T/t=0.1$, we obtain the same result as the exact one even for $m=50$. At a higher temperature $T/t=2$, however, agreement is less satisfactory, since the number of $m$ is not enough. This result again demonstrates that this DMRG technique works well, in particular, at low temperatures. We note that the position of the lowest-energy peak dose not change with increasing $T$, which is a consequence of the spin-charge separation.~\cite{onodera}

\begin{figure}[t]
\begin{center}
\includegraphics[width=90mm]{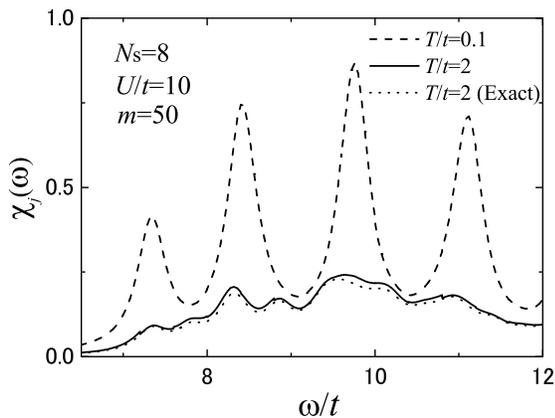}
\caption{Dynamical current-current correlation functions in the eight-site Hubbard chain at half filling with $U/t=10$. $M=16$.
}
\end{center}
\label{fig3}
\end{figure}

We have shown that it is necessary to perform random samplings of the state $\left|\xi\right>$ to calculate physical quantities in the process of the measurement in DMRG. The number of $M$ necessary to get small statistical error is denoted in the captions of each figure. We find that $M$ is dependent on physical quantities. We, thus, need to check an adequate $M$ for each quantity by examining the magnitude of the error bar.
Such a sampling procedure is closely related to that used in the finite-temperature Lanczos method.~\cite{Prelovsek,ltlm} 

As compared with another targeting scheme of finite-temperature DMRG,~\cite{Moukouri} the present method has an advantage that one does not need to divide the Hilbert space with respect to, the $z$ component of total spin, but can treat full of the Hilbert space at once. This reduces a tedious procedure of numerical simulations significantly. 
Furthermore, since this technique is of a simple extension of the zero temperature DMRG supplemented by the polynomial expansion and random sampling, momentum-dependent quantities can also be calculated unlike the case of the transfer-matrix DMRG. However, it is not practical to use the present method at high temperatures, since the large $m$ is required. We also note that, although there is no restriction in principle to apply this technique to complicated models with long-range interactions, a large number of $m$ is required even at low temperatures as is the case of the zero-temperature DMRG.

In summary, 
a different DMRG technique has been developed in order to calculate both static and dynamical quantities at low temperatures. 
This technique is of a straightforward extension of a single-target DMRG procedure, except that the target state is evaluated by the regulated polynomial expansion and a random-sampling and averaging procedure are employed for the measurement of physical quantities.
By using the proposed technique, static and dynamical quantities in the 1D half-filled Hubbard chains have been calculated, and it has been demonstrated that the technique works well at low temperatures.
This technique would be useful as one of the DMRG techniques at low temperatures.

 The authors thank H. Matsueda, M. Itoh, T. Mutou, P. Prelov\v{s}ek, and S. Maekawa for useful discussions. 
This work was supported by Next Generation Supercomputing Project of Nanoscience Program, Grant-in-Aid for Scientific Research from MEXT, and the Academic Center for Computing and Media Studies, Kyoto University (ACCMS) for the use of the computing facilities. The numerical calculations were carried out at YITP and ACCMS, Kyoto University, and ISSP, University of Tokyo. This work is also supported in part by the Yukawa International Program for Quark-Hadron Sciences at YITP.



\begin{thebibliography}{99}

\bibitem{White} S. R. White, Phys. Rev. Lett. \textbf{69}, 2863 (1992); S. R. White, Phys. Rev. B \textbf{48}, 10345 (1993).

\bibitem{Jakelmann} E. Jeckelmann, F. Gebhard, and F. H. L. Essler, Phys. Rev. Lett. \textbf{85}, 3910 (2000).

\bibitem{Moukouri} S. Moukouri and L. G. Caron, Phys. Rev. Lett. \textbf{77}, 4640 (1996).

\bibitem{Shibata} N. Shibata, J. Phys. Soc. Jpn. \textbf{66}, 2221 (1997).

\bibitem{Wang} X. Wang and T. Xiang, Phys. Rev. B \textbf{56}, 5061 (1997).

\bibitem{Feiguin} A. E. Feiguin and S. R. White, Phys. Rev. B \textbf{72}, 220401(R) (2005).

\bibitem{Vorter} R. N. Silver and H. R\"oder, Int. J. Mod. Phys. C \textbf{5}, 735 
(1994); R. N. Silver, H. R\"oder, A. F. Voter, and J. D.
Kress, J. Comput. Phys. \textbf{124}, 115 (1996).

\bibitem{LWang} L. W. Wang, Phys. Rev. B \textbf{49}, 10154 (1994); L. W. Wang
and A. Zunger, Phys. Rev. Lett. \textbf{73}, 1039 (1994).

\bibitem{Sota} S. Sota and M. Itoh, J. Phys. Soc. Jpn. \textbf{76}, 054004 (2007).

\bibitem{Prelovsek} J. Jakli\v{c} and P. Prelov\v{s}ek, Phys. Rev. B \textbf{49}, 5065 (1994); J. Jakli\v{c} and P. Prelov\v{s}ek, Adv. Phys. \textbf{49}, 1 (2000).

\bibitem{Matsueda} H. Matsueda, T. Tohyama, and S. Maekawa, Phys. Rev. B \textbf{70}, 033102 (2004).

\bibitem{Deguchi} T. Deguchi, F. H. L. Essler, F. G{\"o}hnann, A. Kl{\"u}mper, V. E. Kopepin, and K. Kusakabe, Phys. Rep. \textbf{331}, 197 (2000).

\bibitem{onodera} H. Onodera, T. Tohyama, and S. Maekawa, Phys. Rev. B \textbf{69}, 245117 (2004).

\bibitem{ltlm} M. Aichhorn, M. Daghofer, H. G. Evertz, and W. von der Linden, Phys. Rev. B \textbf{67}, 161103(R) (2003).

\end{thebibliography}
\end{document}